\documentclass[aip,jcp,preprint,noshowkeys,superscriptaddress]{revtex4-2}
\usepackage{graphicx,dcolumn,bm,xcolor,microtype,multirow,amscd,amsmath,amssymb,amsfonts,physics,wrapfig,bbold,siunitx,xspace,braket,txfonts}
\usepackage[version=4]{mhchem}

\usepackage[utf8]{inputenc}
\usepackage[T1]{fontenc}
\usepackage{longtable}
\usepackage{threeparttable}
\usepackage{multirow}
\usepackage{calrsfs}
\usepackage{tikz}
\usepackage{chemfig}
\usepackage{mhchem}
\usepackage{placeins}
\usepackage{hyperref}
\hypersetup{
    colorlinks,
    linkcolor={red!50!black},
    citecolor={red!70!black},
    urlcolor={red!80!black}
}

\usepackage{listings}
\definecolor{codegreen}{rgb}{0.58,0.4,0.2}
\definecolor{codegray}{rgb}{0.5,0.5,0.5}
\definecolor{codepurple}{rgb}{0.25,0.35,0.55}
\definecolor{codeblue}{rgb}{0.30,0.60,0.8}
\definecolor{backcolour}{rgb}{0.98,0.98,0.98}
\definecolor{mygray}{rgb}{0.5,0.5,0.5}

\definecolor{sqred}{rgb}{0.85,0.1,0.1}
\definecolor{sqgreen}{rgb}{0.25,0.65,0.15}
\definecolor{sqorange}{rgb}{0.90,0.50,0.15}
\definecolor{sqblue}{rgb}{0.10,0.3,0.60}

\lstdefinestyle{mystyle}{
    backgroundcolor=\color{backcolour},
    commentstyle=\color{codegreen},
    keywordstyle=\color{codeblue},
    numberstyle=\tiny\color{codegray},
    stringstyle=\color{codepurple},
    basicstyle=\ttfamily\footnotesize,
    breakatwhitespace=false,
    breaklines=true,
    captionpos=b,
    keepspaces=true,
    numbers=left,
    numbersep=5pt,
    numberstyle=\ttfamily\tiny\color{mygray},
    showspaces=false,
    showstringspaces=false,
    showtabs=false,
    tabsize=2
  }

  \newcolumntype{d}{D{.}{.}{-1}}

\newcommand{\cre}[1]{\ensuremath{a^\dagger_{#1}}}
\newcommand{\ani}[1]{\ensuremath{a_{#1}}}

\newcommand{\fnt}{\footnotetext}
\newcommand{\fnm}{\footnotemark}
\newcommand{\centpair}[2]{[#1, #2]}

\newcommand{\irrep}[1]{\ensuremath{\Gamma^{#1}}}
\newcommand{\irrepof}[1]{\ensuremath{\Gamma_{#1}}}

\newcommand{\LCPQ}{Laboratoire de Chimie et Physique Quantiques (UMR 5626), Universit\'e de Toulouse, CNRS, Bat. 3R1b4, 118 route de Narbonne, 31062, Toulouse Cedex 09, France}
\newcommand{\UDS}{Physikalische und Theoretische Chemie, Universität des Saarlandes, Campus, Geb. B2.2, 66123,  Saarbrücken, Germany}
\newcommand{\JGU}{Department Chemie, Johannes Gutenberg-Universit\"at Mainz, Duesbergweg 10-14, 55128,  Mainz, Germany}
\newcommand{\Hyl}{Hylleraas Centre for Quantum Molecular Sciences, Department of Chemistry, University of Oslo, P.O. Box 1033 Blindern, 0315, Oslo, Norway}

\begin{document}	

\title{Exploitation of complex Abelian point groups in quantum-chemical calculations}

\author{Marios-Petros \surname{Kitsaras}}
	\email{kitsaras@irsamc.ups-tlse.fr}
	\affiliation{\LCPQ}
    \affiliation{\UDS}
	\affiliation{\JGU}

\author{Stella Stopkowicz}
	\email{stella.stopkowicz@uni-saarland.de}
	\affiliation{\UDS}
    \affiliation{\Hyl}
	\affiliation{\JGU}
	
\begin{abstract}
Quantum-chemical calculations often make use of point-group theory to exploit molecular symmetry, resulting in a reduction of the computational cost and in insights into the electronic structure. This exploitation is often limited to subgroups of $D_{2h}$ which are Abelian with real characters. 
Here, we extend the symmetry exploitation to Abelian point groups with complex characters. 
Such point groups are often encountered in calculations that involve finite magnetic fields, though their occurrence is not limited to these cases alone.
We present the evaluation of integrals over symmetry-adapted orbitals using the double-coset decomposition, as well as the use of these symmetries in the contractions needed within post Hartree Fock calculations in the context of block tensors. 
Efficiency gains are discussed for four simple hydrocarbons that exhibit a complex Abelian point group in the presence of a magnetic field.
\end{abstract}

\maketitle

\section{Introduction}

Symmetry is a fundamental organizing principle in chemistry because it rigorously constrains how molecular structures and electronic states transform under geometric operations. It provides a powerful framework to interpret molecular properties, such as the classification of vibrational modes in spectroscopy, the assignment of molecular orbitals, and the prediction of allowed electronic transitions.
As an example, standard texts such as 
Chemical Applications of Group Theory (Cotton, Wiley)\cite{Cotton1971}
illustrate how group theory links molecular structure, observables, and physical behavior.

In quantum-chemical computations, symmetry is exploited both to interpret electronic structure and to reduce computational cost in terms of both memory requirements as well as computation time.\cite{Davidson1975,Dupuis1977,Dupuis1978,Taylor1985,Taylor1986,Stanton1991a,Gauss1991,Haeser1991,Pausch2021,Nottoli2023,Huynh2024,Melega2025} 
The central principle is that (approximate) solutions to the  Schrödinger equation transform like one of the irreducible representation (IRREP) of the molecular point group (PG). In this way, due to resulting selection rules, the computation and handling of integrals, amplitudes, and intermediate quantities that vanish due to symmetry can be avoided.
Most often such symmetry implementations are based on Abelian PGs with real-valued characters as wavefunctions of interest are formulated in a purely real manner in the majority of quantum-chemical calculations.
However, there are situations in which real representations are insufficient, motivating the use of Abelian PGs with complex-valued characters to fully exploit molecular symmetry and reduce computational cost.
This is true in particular in cases where the wavefunction of the system under consideration becomes complex. 
Complex wavefunctions occur, for example, for molecules in finite magnetic fields\cite{Tellgren2008}  or in relativistic calculations that consider spin-orbit effects in a non-perturbative manner.\cite{Saue2011} 
But even in standard field-free non-relativistic calculations, complex wavefunctions can arise. 
This includes, for example, open-shell $\Pi$, $\Delta$, etc. states as eigenvalues of the angular momentum operator, as well as
non-Hermitian parameterizations of the Hamiltonian, like in equation-of-motion coupled cluster (EOM-CC) theory, when treating accidentally degenerate excited states of molecules whose symmetry is described by a complex Abelian PG (e.g., excited states of B(OH)$_3$, $C_{3h}$).\cite{Grazioli2025}\textsuperscript{,}\footnote{In Hermitian parameterizations this issue is circumvented by forming real wavefunctions as linear combinations of the \emph{exactly} degenerate complex solutions. Note that this issue in EOM-CC theory is directly related to the known problem known problem for treating conical intersections between states of the same symmetry within CC theory.\cite{Koehn2007} In such cases, a complex wavefunction cannot be avoided even when a real Abelian subgroup is employed.}

In general, symmetry exploitation in quantum-chemical programs occurs throughout the calculations, starting from the level of integral evaluation up to the tensor-contraction level involved in electronic-structure approximations (self-consistent field (SCF), post-Hartree-Fock (HF), etc.).
At the level of integral evaluation, symmetry can be used to reduce the number of integrals that need to be computed. Two widely used approaches are the petite-list method\cite{Dupuis1977,Dupuis1978} and the double-coset decomposition (DCD).\cite{Davidson1975} Both function by identifying the symmetry-equivalent combinations of basis functions and avoid repeated evaluations of symmetry-equivalent integrals. While the petite-list approach is formally and algorithmically simpler without any drawback in the case of non-Abelian groups, the resulting integral matrices are not symmetry-adapted and the rule of vanishing integrals cannot be directly exploited. The DCD on the other hand leads to a block structure of the integral matrices as it implicitly symmetry adapts the basis functions. 
While a formulation of the DCD in the case of non-Abelian PGs has 
been derived, it relies on the full matrices of the IRREPs 
rather than their characters, which makes its practical implementation more involved. 
For HF and post-HF electron correlation treatments, all quantities are typically expressed in terms of symmetry-adapted MOs. In this context, the direct-product decomposition (DPD) scheme\cite{Stanton1991a} enables the systematic symmetry blocking of amplitudes and intermediates, allowing efficient exploitation of Abelian PGs.
Moreover, recent work has extended symmetry exploitation using the DPD scheme to  non‑Abelian PGs, demonstrating potential computational savings in many‑body methods when such symmetries are fully utilized.\cite{Hellmann2026}

In this work, we focus on the case of molecules in a finite magnetic field. 
Symmetry handling for Abelian PGs with real characters in the \textsc{CFOUR} program\cite{Matthews_2020,cfour} can be traced back to the corresponding implementation in the \textsc{aces II}\cite{aces2,Almloef1997} program package. 
In the more recent \texttt{mint}\cite{mint} integral package within \textsc{cfour}, symmetry exploitation was implemented by Gauss for the evaluation of integrals over both standard Gaussian basis functions and London orbitals\cite{London1937}. In the context of finite magnetic fields, this symmetry handling was further extended to the HF–SCF procedure itself.
For subsequent post-HF calculations, a corresponding symmetry handling was introduced within the \textsc{qcumbre}\cite{Hampe2017,qcumbre} program.\cite{Kitsaras2023}
These developments have enabled the symmetry exploitation at the finite-field (ff) HF, second-order M{\o}ller-Plesset perturbation theory, and various flavors of Coupled-Cluster (CC) and Equation-of-Motion (EOM) CC levels of theory.\cite{Hampe2017,Hampe2019,Hampe2020,Blaschke2021,Kitsaras2023,Kitsaras2024,Kitsaras2025,Grazioli2025,Monzel2025}
Here, we extend this symmetry handling to account also for Abelian PGs with complex characters as such PGs often describe the molecular symmetry in the presence of a magnetic field.\cite{Pausch2021} 
We note in this context that, recently, tools for symbolic symmetry analysis specialized for magnetic PGs have been developed.\cite{Huynh2024}\textsuperscript{,}\footnote{Magnetic PGs are an extension of standard molecular PGs that account for both unitary and antiunitary symmetry operations.}

We present an implementation of Abelian PGs with complex characters 
in a quantum-chemical framework. We first revisit the DCD for  such groups and derive working equations suitable for a practical integral evaluation. 
Building on this, we present the theoretical background for the respective exploitation of symmetry 
in HF and post-HF correlation methods, including CC, EOM-CC, and the evaluation of molecular properties. 
Lastly, we report on the resulting computational savings and performance improvements, illustrated by ground state CCSD energies.

\section{Theory}

Complex Abelian PGs possess all the properties of real Abelian PGs  that make them attractive for their exploitation within quantum-chemical calculations. They only have one-dimensional IRREPs, and their symmetry elements commute. However, they have complex entries in their character tables and their matrix respresentations,\cite{Altmann2011} which hinders their exploitation, as many quantum-chemical program packages are restricted to real numbers. IRREPs of a group with complex entries come in complex-conjugate pairs. This practically means
that for each IRREP with complex matrix representations, $\irrep{\alpha}$, there exist an IRREP \irrep{\alpha *} whose matrices are the complex conjugate of each other without them being equivalent.
The direct product between this complex-conjugate pair results in the totally symmetric IRREP
\begin{equation} 
\irrep\alpha\otimes\irrep{\alpha *} = \irrep1.
\end{equation}

In this work, the theoretical framework for the exploitation of complex Abelian PGs in quantum-chemical calculations is presented. It encompasses both the employment of these PGs in the calculation of integrals over atomic basis functions as well as the symmetry handling in SCF and post-HF approaches. Key differences and additional considerations with respect to the standard real Abelian PGs, i.e., $D_{2h}$ and its subgroups, will be discussed.  

\subsection{Double-coset decomposition in complex Abelian point groups \label{sec:DCD}}
The DCD for the calculation of integrals of atomic orbitals has been first presented by \citeauthor{Davidson1975}\cite{Davidson1975} and has later been extended to integral derivatives.\cite{Taylor1985,Taylor1986} 
Closely related developments in the exploitation of symmetry for integral evaluation were implemented earlier and reported by \citeauthor{Almloef1997}.\cite{Almloef1997}
We  refer the reader to ref.~\onlinecite{ESQC2019} for a pedagogical introduction to the DCD. Here, we limit the presentation to the specific case of complex Abelian PGs. 
For clarity, the notation used in this manuscript for point-group theory is summarized in Tab.~\ref{tab:sym_not}.

\renewcommand{\arraystretch}{1.5}
\begin{table*}[t]
    \centering
    \begin{tabular}{c|l}
        Notation & Use \\ \hline
        $G_i$ & {\raggedright\arraybackslash Symmetry element} \\
    $\hat{G_i}$ &  \raggedright\arraybackslash Symmetry operation \\
    $\mathcal{G}$ &  \raggedright\arraybackslash Point group \\
    $\irrep\alpha$ &  \raggedright\arraybackslash IRREP with index $\alpha$ \\
    $\gamma^\alpha_{G_i}$ &  \raggedright\arraybackslash Character of IRREP $\alpha$ and symmetry element $G_i$ \\
    $n_G$ & {\raggedright\arraybackslash Number of elements within group $\mathcal{G}$} \\
    $R$ & A specific choice of $\mathcal{F}$ and $\mathcal{H}$ subgroups for the construction of double cosets $\mathcal{F}G_i\mathcal{H}$ \\
    $R_k$ & Double-coset representatives that generate all distinct double cosets for $R$ \\
    $n^R_\mathrm{DCR}$ &  Number of distinct double cosets for $R$  \\
    $\lambda_R$ & {\raggedright\arraybackslash Degeneracy of the double cosets generated for $R$} \\
    $\chi_{\mu,A}$ & {\raggedright\arraybackslash Function in the AO set with index $\mu$ and atomic center $A$} \\
    $X^\alpha_{\mu,A}$ & {\raggedright\arraybackslash Function in the SAO set with index $\mu$, delocalized over atomic centers of type $A$, transforms as IRREP $\irrep\alpha$} \\
    $G_i(A)$ & {\raggedright\arraybackslash Symmetry-equivalent center to $A$, into which $A$ transforms after applying symmetry operation $\hat G_i$} \\
    \centpair{$A$}{$B$} & A pair of centers $A$ and $B$\\
    $\mathcal{U}$ & Stabilizer of center $A$ \\
    $\irrepof{p}$ &  \raggedright\arraybackslash IRREP of symmetry-adapted object $p$ \\
    \end{tabular}   
    \caption{Collection of the different notations used within the masnuscript.}
    \label{tab:sym_not}
\end{table*}
\renewcommand{\arraystretch}{1.}

In this work, the calculation of one- and two-electron integrals takes place in the AO basis $\{ \chi_{\mu,A} \}$. The index $\mu$ enumerates the AOs centered at nucleus $A$. 
However, since the AOs do not necessarily transform according to the IRREPs of the PG of the molecule, it is useful to work with symmetry-adapted atomic orbitals (SAOs) $\{ X_{\mu,A}^\alpha\}$.
SAOs are enumerated by index $\mu$ as well, and transform according to IRREP $\irrep\alpha$. Furthermore, they are typically delocalized over all symmetry-related sites obtained by applying the  symmetry operations of the group to center $A$. 
Given a complex Abelian PG $\mathcal{G}$ with symmetry elements $\{G_i\}$, a transformation from the AO to the SAO set can be defined as
\begin{align}
    \hat{\mathcal{P}}^\alpha \chi_{\mu,A} = \frac{1}{n_G} \sum_i^{n_G}{\gamma^{\alpha *}_{G_i}}\hat{G}_i\;\chi_{\mu,A} = X^\alpha_{\mu,A}, \label{eq:proj}
\end{align}
using the projection operator $\hat{\mathcal{P}}^\alpha$.
Here,  $\gamma^{\alpha}_{G_i}$ is the character of symmetry element $G_i$ in the IRREP $\irrep\alpha$ and $\hat{G}_i$ is the operator associated with the corresponding symmetry element. 
Note that unlike in the case of real Abelian groups, the  complex conjugation for $\gamma^{\alpha}_{G_i}$ cannot be omitted. 
A drawback of an explicit transformation such as that in Eq.~\eqref{eq:proj} is that the summation over all $n_G$ elements of the group typically leads to redundant contributions, i.e., the same functions are generated multiple times. 
Rather than proceeding in this manner and subsequently factorizing the terms in the final expression, the DCD is employed to eliminate these redundancies straight from the beginning, as explained below. 

A double coset forms a subset of the elements of the group without being a group itself. 
This disjoint partitioning takes the form
\begin{equation}
    \mathcal{F}G_i\mathcal{H} = \{F_1G_iH_1, F_1G_iH_2, \dots, F_2G_iH_1, \dots \} ,
    \label{eq:disjoint_partitioning}
\end{equation} 
where $\mathcal{F}$ and $\mathcal{H}$ are subgroups of $\mathcal{G}$. A given choice of subgroups is denoted by $R$. 
Each element of a double coset exhibits a degeneracy $\lambda_R$, meaning that it appears $\lambda_R$ times within the set. All elements within a double coset have the same degeneracy. 
Furthermore, for an Abelian group, this degeneracy coincides with the number of common elements between the two sets:
\begin{equation}
    \lambda_R = |\mathcal{F}\cap \mathcal{H} |.
\end{equation} 
As such, for a choice $R$, all double cosets have the same degeneracy.
Different choices of the central symmetry element $G_i$ of a given PG 
in Eq.~\eqref{eq:disjoint_partitioning} may result in either distinct (no elements in common) or equivalent (all elements in common) sets. 
In order to systematically characterize the distinct sets, one arbitrary element $R_k$ is selected 
from within each distinct double coset. 
These elements $R_k$ are denoted as double-coset representatives (DCR) for a specific choice of subgroups $\mathcal{F}$ and  $\mathcal{H}$, with $n_\mathrm{DCR}^R$ the number of distinct cosets.
The selected DCRs can then be used to rewrite Eq.~\eqref{eq:proj}, thereby replacing $\hat{G}_i$ by $\hat{F}_i \hat{R}_k \hat{H}_j$. 
Furthermore, the redundant sum over all group elements $\sum_i^{n_G} $  in Eq.~\eqref{eq:proj} can be 
decomposed into the sum over elements of the subgroups $\sum_{i}^{n_F}$ and $ \sum_{j}^{n_H}$, as well as the DCRs $\sum_k^{n_\mathrm{DCR}^R}$.  
The resulting decomposed expression for the projection, i.e., the DCD, is
\begin{align}
\hat{\mathcal{P}}^\alpha \chi_{\mu,A} =
    \frac{1}{n_G \lambda_{R}} 
    \sum_i^{n_F}
    \sum_j^{n_H}
    \sum_{k}^{n_\mathrm{DCR}^R}
    \gamma^{\alpha *}_{F_i}
    \gamma^{\alpha *}_{R_k}
    \gamma^{\alpha *}_{H_j}
    \hat{F}_i
    \hat{R}_k
    \hat{H}_j\;
    \chi_{\mu,A}. \label{eq:DCD}
\end{align}  
The appropriate choice of the left and right subgroups is the key to achieving the desired factorization, as it allows the elimination of the corresponding sums over elements of the subgroups $\mathcal{F}$ and $\mathcal{H}$, i.e., the sums over $i$ and $j$ in Eq.~\eqref{eq:DCD}. 
It is stressed, that the remaining sum over $k$ has fewer elements than the original sum in Eq.~\eqref{eq:proj}.

Next, it will be discussed how to arrive at a suitable choice for $\mathcal{F}$ and $\mathcal{H}$, thereby introducing the concept of stabilizers.  
Let us first consider the action of the symmetry operation $\hat{G}_i$. By definition, its action on a function $f$ of the 3D space vector $\mathbf{r}$ is 
\begin{align}
    \hat{G}_i f(\mathbf{r}) = f(\hat{G}_i^{-1}\mathbf{r}),
\end{align}
with $G_i^{-1}$ as the inverse element of $G_i$ acting on $\mathbf r$. 
Hence, the action of  $\hat{G}_i$ on an AO leads to a linear combination of functions from the same set
\begin{align}
    \hat{G}_i\chi_{\mu,A} = \sum_{\mu'} C^A_{\mu \mu'}(G_i)\; \chi_{\mu',G_i(A)}, \label{eq:rot_AO}
\end{align}
where $G_i(A)$ is the center into which $A$ transforms after application of the symmetry operation $G_i$. The coefficient matrices $C^A_{\mu \mu'}(G_i)$ depend on the angular momenta of the basis functions and differ depending on whether Cartesian or spherical basis sets are used, respectively. In the latter case, for example, the coefficient matrices coincide with the Wigner-D matrices.\cite{Wigner1959}

In the general case, as seen in Eq.~\eqref{eq:rot_AO}, the symmetry operation moves the original center of the AO function $A$ to a \textit{different} symmetry-equivalent center $G_i(A)$. \textit{Some} symmetry operations, however, leave specific atomic centers invariant. More concretely, the set of all symmetry elements $U_i$ of the full group $\mathcal{G}$ that leave a given atomic center $A$ invariant 

form a subgroup $\mathcal{U}$, which is referred to as the stabilizer of $A$. 
Thus, the choice of subgroups $\mathcal{F}$ and $\mathcal{H}$ for the DCD in Eq.~\eqref{eq:DCD} can be based on the stabilizers of the respective centers in order to precalculate the factorization mentioned above. 

Having established the guiding principles underlying the DCD of the projection operator in Eq.~\eqref{eq:DCD}, the elimination of redundant contributions in the integral calculation over SAOs is addressed in the following paragraphs. 
Note that the AO to SAO transformation is thereby performed implicitly. 

A one-electron integral over SAOs 
   $ \bra{X^\alpha_{\mu,A}} \hat{O} \ket{X^\beta_{\nu,B}}$
can be rewritten using the DCD in Eq.~\eqref{eq:DCD}
as
\begin{equation}
\begin{gathered}
 \bra{X^\alpha_{\mu,A}} \hat{O} \ket{X^\beta_{\nu,B}} =\\   
 \frac{1}{\lambda_R}\sum_\gamma I_{\alpha \gamma \beta} \sum_{\mu'}  {\Lambda^{\alpha A*}_{\mu\mu'}(E)} \sum_{j}^{n_\mathrm{DCR}^R}\sum_{\nu'} \Lambda^{\beta B}_{\nu\nu'}(R_j)  \bra{\chi_{\mu',A}}\hat{O}^\gamma \ket{\chi_{\nu',R_j(B)}},  \label{eq:oneDCR_nosym}
\end{gathered}
\end{equation}
where $E$ is the identity symmetry operation.
Note that in the above expression, the subgroups $\mathcal{F}$ and $\mathcal{H}$ of the double coset have been chosen as the stabilizers of $A$ and $B$, respectively.
The matrices ${\Lambda^{\alpha A}_{\mu\mu'}(G_i)}$ contain
the corresponding sums over the elements of the subgroups 
\begin{align}
    \Lambda^{\alpha A}_{\mu\mu'}(G_i) &= \frac{1}{n_G} \sum_k^{n_U} \gamma^{\alpha*}_{G_iU_k} C^A_{\mu\mu'}(G_iU_k).  \label{eq:lambda_sym_dcd}
\end{align}
Note that in Eq.~\eqref{eq:lambda_sym_dcd}, the sum over $k$ involves the elements $U_k$ of the subgroup $\mathcal{U}$, i.e., the stabilizer of $A$. 
The operator $\hat{O}^\gamma$ in Eq.~\eqref{eq:oneDCR_nosym} is the result of a symmetry projection $\hat{\mathcal{P}}^\gamma\hat{O}$ of the original operator onto IRREP $\irrep{\gamma}$. 
Thus, the sum over $\gamma$ in Eq.~\eqref{eq:oneDCR_nosym} runs over all non-vanishing projections. 
Lastly, the prefactor $I_{\alpha \gamma \beta}$  acts as a selection rule: it equals $n_G$ if the direct product $\irrep{\alpha} \otimes \irrep{\gamma*} \otimes \irrep{\beta*}$ corresponds to the totally symmetric IRREP $\irrep1$, and is zero otherwise.

As mentioned earlier, choosing the left and right subgroups of the double coset as the stabilizers of $A$ and $B$ results in pre-calculating the final factorization of the redundant terms of the projection in advance. It can hence be recognized that the sum over the DCRs $R_j$ only affects the ket of the integral, transforming the center of the respective AO function from $B$ to $R_j(B)$.  Through this restricted sum, the minimum number of non-redundant  \centpair{$A$}{$R_j(B)$} pairs needed for the calculation of the one-electron integral are generated.

Similarly to the one-electron case, the two-electron integrals over SAOs 
expressed in Mulliken notation, can be rewritten as
    \begin{equation}
    \begin{gathered}
 (X^\alpha_{\mu,A} X^\beta_{\nu,B} | X^\gamma_{\rho,C} X^\delta_{\sigma,D})=\\
   \frac{I_{\alpha \beta \gamma \delta} }{\lambda_T}\sum_{\mu'\nu'\rho'\sigma'}{\Lambda^{\alpha A*}_{\mu\mu'}(E)} 
   \sum_{j}^{n_\mathrm{DCR}^R} \Lambda^{\beta B}_{\nu\nu'}(R_j) 
   \sum_{l}^{n_\mathrm{DCR}^T} {\Lambda^{\gamma C*}_{\rho\rho'}}(T_l) 
   \sum_{k}^{n_\mathrm{DCR}^S} \Lambda^{\delta D}_{\sigma\sigma'}(T_lS_k)  \cdot \\
   (   \chi_{\mu',A}    \chi_{\nu',R_j(B)} |    \chi_{\rho',T_l(C)}    \chi_{\sigma',T_lS_k(D)}) . \label{eq:twoint_final}
 \end{gathered}
\end{equation}
Here, $R_j$ are the DCRs of the double cosets with the stabilizers of $A$ and $B$, $S_k$ are the DCRs of the double cosets with stabilizers of $C$ and $D$.  $T_l$, on the other hand, are the DCRs of the double cosets with stabilizers of the center pairs \centpair{$A$}{$R_j(B)$} and \centpair{$C$}{$S_k(D)$}. The prefactor $I_{\alpha \beta \gamma \delta}$ acts as a selection rule associated with the direct product $\irrep\alpha \otimes \irrep{\beta*} \otimes \irrep\gamma \otimes \irrep{\delta*}$. Again, the non-redundant combinations of centers $A$, $B$, $C$, and $D$ necessary for the calculation of the two-electron integrals are generated by the DCRs through the DCD.

A detailed derivation of the expressions in Eqs.~\eqref{eq:oneDCR_nosym} and \eqref{eq:twoint_final} can be found in ref.~\onlinecite{Kitsaras2023}.

The main difference between the simplification from the general case to real or complex Abelian PGs, apart from the presence of complex-valued characters, lies in Eq.~\eqref{eq:rot_AO}: 
In the case of real Abelian PGs, the coefficient matrix $C^A_{\mu \mu'}(G_i)$ always has one single non-vanishing element and can be simplified to a parity prefactor $\pm 1$.\cite{Davidson1975,ESQC2019} 
Hence, the expressions for the one-  and two-  electron integrals simplify to  
\begin{gather}
\begin{gathered}
    \bra{X^\alpha_{\mu,A}} \hat{O} \ket{X^\beta_{\nu,B}} =\\ \frac{1}{\lambda_R}\sum_\gamma I_{\alpha \gamma \beta}   {\Lambda^{\alpha A}_{\mu\mu}(E)} \sum_{j}^{n_\mathrm{DCR}^R} \Lambda^{\beta B}_{\nu\nu}(R_j)  \bra{\chi_{\mu,A}}\hat{O}^\gamma \ket{\chi_{\nu,R_j(B)}},  \label{eq:oneDCR_nosym_re}
\end{gathered}
\end{gather}
and
\begin{gather}
   \begin{gathered}
   (X^\alpha_{\mu,A} X^\beta_{\nu,B} | X^\gamma_{\rho,C} X^\delta_{\sigma,D})=\\
   \frac{I_{\alpha \beta \gamma \delta} }{\lambda_T}{\Lambda^{\alpha A}_{\mu\mu}(E)} 
   \sum_{j}^{n_\mathrm{DCR}^R} \Lambda^{\beta B}_{\nu\nu}(R_j) 
   \sum_{l}^{n_\mathrm{DCR}^T} {\Lambda^{\gamma C}_{\rho\rho}}(T_l) 
   \sum_{k}^{n_\mathrm{DCR}^S} \Lambda^{\delta D}_{\sigma\sigma}(T_lS_k)  \cdot \\
   (   \chi_{\mu,A}    \chi_{\nu,R_j(B)} |    \chi_{\rho,T_l(C)}    \chi_{\sigma,T_lS_k(D)})  \label{eq:twoint_final_re}
   \end{gathered}
\end{gather}
in the case of real Abelian PGs, respectively.
In the case of complex Abelian PGs, however, the coefficient matrices possess more than one non-vanishing element. 
Specifically, the matrices of proper and improper rotations around the principal rotation axis which are present in these groups create a linear combination of AOs with the same total-angular momentum quantum number $l$ for a given Gaussian exponent. The number of contributing functions in this linear combination is $(l+1)(l+2)/2$ for a Cartesian basis and $2l+1$ for a spherical basis. In the latter case, this sum is further limited to a maximum of two contributions from AOs with the same absolute 
magnetic quantum number $|m_l|$ associated with angular momentum $l$,  
since only rotations over the principal rotation axis exist in complex Abelian PGs.  As such, the sums over AOs in Eqs.~\eqref{eq:oneDCR_nosym} and \eqref{eq:twoint_final} that arise from Eq.~\eqref{eq:rot_AO} do not simplify to one single contribution. Yet, they run over a limited number of AOs that can be easily predetermined.\footnote{Further details on how to calculate the elements of the coefficient matrices $C^A_{\mu \mu'}(G_i)$ for each symmetry elements in the case of a Cartesian or a spherical basis set can be found in ref.~\onlinecite{Kitsaras2023}.}

\subsection{Complex Abelian point groups in second quantization}
In this section, the symmetry properties of the targeted wavefunction in the case of complex Abelian PGs are examined within the second-quantization formalism. Note that, in contrast to real Abelian PGs, proper care must be taken to account for the complex conjugation of the matrix representations of the IRREPs.
The discussion addresses the treatment of HF, CC, and EOM-CC approximations, 
as well as the evaluation of properties via the construction of density matrices.

In second quantization, a particle creation operator with respect to the true vacuum, i.e., $\cre{p}$ can be associated with the $\irrepof{p}$ IRREP of the symmetry-adapted molecular orbital and an annihilator $\ani{p}$ with the IRREP corresponding to the complex conjugate matrices  $\irrepof{p}^*$. 
The IRREP of the Fermi vacuum, 
\begin{align}
    \ket{\Phi_0} = \prod_i^\mathrm{occ} \cre{i}\ket{\mathrm{vac}},
\end{align} 
which in this discussion corresponds to the HF state, 
is determined by the direct product of the IRREPs of the occupied ($i,\ j,\ k, \dots$) orbitals \begin{equation}
    \irrepof{\Phi_0}=\irrepof{i}\otimes\irrepof{j} \otimes \dots
\end{equation}
In practical terms, one can select the IRREP of the 
HF state 
by choosing the number of occupied orbitals from each IRREP.

For an arbitrary operator expressed in second quantization
\begin{equation}
    \hat{O}= \sum_{pq\dots rs \dots}o^{pq...}_{rs...}\cre{p}\cre{q} \dots \ani{s} \ani{r},
\end{equation}
each amplitude $o^{pq...}_{rs...}$ is associated with IRREP
\begin{equation}
    \irrepof{p}\otimes\irrepof{q} \otimes ... \otimes (\irrepof{r})^*\otimes (\irrepof{s})^* \otimes ... =\irrepof{o^{pq...}_{rs...}}
\end{equation}
If this operator is symmetry-adapted and transforms as $\irrepof{\hat{O}}$, all amplitudes for which
\begin{align}
    \irrepof{o^{pq...}_{rs...}} \neq \irrepof{\hat{O}} \label{eq:o_select}
\end{align}
are zero and Eq.~\ref{eq:o_select} thus acts as a selection rule.

It is noted that the electronic Hamiltonian  $\hat{H}$ and all its individual contributions transform as the totally symmetric IRREP $\irrepof{\hat{H}}=\irrep1$. 
Furthermore, the operators used to describe electron correlation effects for the ground state by working on 
the HF reference are themselves totally symmetric. 
This follows from the fact that the correlated wavefunction has the same IRREP as the Fermi vacuum $\irrepof{\Phi_0} = \irrepof{\Psi_\mathrm{corr}}$.
One such example is the cluster operator $\hat{T}$ in CC theory,\cite{Shavitt2009} i. e., 
\begin{align}
    \ket{\Psi_\mathrm{corr}} = e^{\hat{T}}\ket{\Phi_0}.
\end{align}

In the EOM-CC approach,\cite{Shavitt2009} an excited state is expressed via an excitation starting from a correlated reference wavefunction 
\begin{align}
    \ket{\Psi_\mathrm{exc}} = \hat{R} \ket{\Psi_\mathrm{corr}}.
\end{align}
The IRREP of $\hat{R}$ can be determined via the IRREP of the targeted excited state  as
\begin{align}
    \irrepof{\hat{R}} = \irrepof{\Psi^{}_\mathrm{exc}} \otimes  \irrepof{\Psi_\mathrm{corr}}^*.
\end{align}
The respective deexcitation operator $\hat{L}$ transforms accordingly as $\irrepof{\hat{L}}^{} = \irrepof{\hat{R}}^*$.   

Lastly, it is noted that density matrices $\mathbf{D}$ typically used for the calculation of properties are necessarily associated with $\irrep1$ irrespective of the IRREP of the electronic wavefunction, as they  arise from expectation-value expressions. Transition-density matrices $\mathbf{D}^{(m)\rightarrow (n)}$ between states $(m)$ and $(n)$, however arise from integrals of the form $\bra{\Psi^{(n)}} \dots \ket{\Psi^{(m)}}$ and transform as the IRREPs of the transition $\irrepof{\mathrm{tr}} = \irrepof{\Psi^{(n)}}^* \otimes \irrepof{\Psi^{(m)}}^{}$.
In the EOM-CC approach $\irrepof{\mathrm{tr}}$ can be determined from the left and right EOM vectors as
\begin{align}
    \irrepof{\mathrm{tr}} = \irrepof{\hat{L}^{(n)}} \otimes \irrepof{\hat{R}^{(m)}}.
\end{align}

\subsection{Handling of symmetry-adapted tensors \label{sec:tens}}
A major computational advantage of symmetry exploitation is the reduction of cost that can be achieved by the selection rule in Eq.~\ref{eq:o_select}. To exploit it, we use the following approach to handle the amplitude tensors: The orbital indices $p,\ q, \dots$ are sorted according to their IRREP. This leads to a block structure of the tensors, where the IRREP of each index within a block is constant. Non-vanishing blocks with index $t$ are given a tag number $t_\mathrm{num}$ and vanishing blocks are ignored. Assuming an $n$-dimensional tensor $O_{p_1, p_2, \dots p_n}$ the tag of each block is 
\begin{align}
    t_\mathrm{num} = \sum_{i=1}^n {(\irrepof{p_i})}_\mathrm{num} n_G^{i-1}. 
\end{align}
Here ${(\irrep\alpha)}_\mathrm{num} = \alpha-1 $ is the enumeration tag of the IRREPs with the totally symmetric IRREP  as the first element ${(\irrep1)}_\mathrm{num}=0$. For given dimensions of the tensor $n$ and the order of the group $n_G$, the tag is unique and directly identifies the IRREP of each index in the block. More importantly, the blocks are ordered in ascending tag number. This sorting allows the utilization of a binary search algorithm with logarithmic scaling, when a specific block needs to be selected. 
By considering only the $n_G^{n-1}$ non-vanishing blocks, the memory requirement is reduced proportional to $\mathcal{O}(n_G)$, i.e., instead of storing $N^n$ elements for the tensor, 
each block has $\left (\frac{N}{n_G}\right )^n$ elements in the case of equally distributed indices, resulting in $n_G^{n-1}\left (\frac{N}{n_G}\right )^n = \frac{N^n}{n_G}$ non-zero elements.  
$N$ is the number of basis functions, which roughly corresponds to the system size.
Accordingly, as resorting 
the entries of the tensors, often also called transposition-steps, scales linearly with the elements of the tensor, transposing a tensor is $\mathcal{O}(n_G)$ more efficient when exploiting the block structure. 

An important operation, which is needed when implementing working equations, is the tensor contraction
\begin{align}
    O_{p_1, p_2, \dots, q_1, q_2, \dots} = \sum_{s_1, s_2, \dots} P_{p_1, p_2, \dots, s_1, s_2, \dots} Q_{s_1, s_2, \dots, q_1, q_2, \dots} \label{eq:tens_contr}
\end{align}
The indices $p_i$ and $q_i$ will be referred to as target indices and $s_i$ as summation indices.
Note that the ordering of indices within the tensors $\mathbf{O}$, $\mathbf{P}$, and $\mathbf{Q}$ is irrelevant for the symmetry handling, but may be important for the actual contraction algorithm, as some algorithms require a transposition step.\cite{Brandejs2026}

Owing to the block structure of the tensors $\mathbf{O}$, $\mathbf{P}$, and $\mathbf{Q}$, the contraction 
is carried out block by block
\begin{align}
    O^t_{p_1, p_2, \dots, q_1, q_2, \dots} =\sum_{\{u,v\}}^{\substack{\textrm{allowed} \\ \textrm{pairs}}} \sum_{s_1, s_2, \dots} P^{u}_{p_1, p_2, \dots, s_1, s_2, \dots} Q^v_{s_1, s_2, \dots, q_1, q_2, \dots}, 
\end{align}
where $t$, $u$, and $v$ are the respective block indices. 
The target indices $p_i$ and $q_i$ in the $u$ and $v$ blocks of tensors $\mathbf{P}$ and $\mathbf{Q}$, respectively,  have to match with the target block $t$ of $\mathbf{O}$. For this reason, the sum over $u$ and $v$ is restricted to these matching blocks. 
Accordingly, not all $u,v$ combinations are allowed since the summation indices $s_i$ have to match as well. The algorithm for this operation thus has the following steps:
\begin{enumerate}
    \item Loop over the $t$ blocks of $\mathbf{O}$.
    \item Preselect the contributing blocks $u$ of $\mathbf{P}$ based on the IRREPs of $p_i$. \label{pres1}
    \item Preselect the contributing blocks $v$ of $\mathbf{Q}$ based on the IRREPs of $q_i$.\label{pres2}
    \item Identify the allowed $\{u,v\}$ pairs from the preselected lists of indices. \label{pairs}
    \item Loop over the allowed pairs $\{u,v\}$ and perform the contraction operation to the individual $\mathbf{P}^u$ and $\mathbf{Q}^v$ blocks, adding their contribution to block $\mathbf{O}^t$ block.  
\end{enumerate}
In this way, the contraction in block structure exhibits savings at the order of $\mathcal{O}(n_G^{2})$.
With  $n_p$, $n_q$, and $n_s$ the number of $p_i$, $q_i$, and $s_i$ indices, respectively,
this means that instead of $N^{n_p+n_q+n_s}$ floating point operations needed for the contraction, $\left (\frac{N}{n_G} \right)^{n_p+n_q+n_s}$ operations are performed for each block combination in the case of equally distributed indices.
In order to find the number of allowed pair combinations $n_\mathrm{pair}$ we note that there are in total $n_G^{n_p+n_q+n_s}$ distinct IRREP combinations for all $p_i$, $q_i$, and $s_i$ indices.
But since the $\{p_i,s_i\}$ and the $\{s_i,q_i\}$ combinations have each been reduced by $n_G$, from the selection rule in the block structures of  $\mathbf{P}$ and $\mathbf{Q}$, the number of allowed pairs is $n_\mathrm{pair}=n_G^{n_p+n_q+n_s-2}$.
As such the total floating point operations in block structure are
\begin{align}
    n_\mathrm{pair}\left (\frac{N}{n_G} \right)^{n_p+n_q+n_s}=\frac{N^{n_p+n_q+n_s}}{n_G^{2}}\;.
\end{align}
This up to quadratic decrease in floating-point operations resulting from symmetry handling is accompanied by the additional preselection steps \ref{pres1}–\ref{pairs} of the algorithm described above.
However, their computational cost is negligible.

In most matrix multiplication routines which are commonly used for tensor contractions, including those in BLAS,\cite{blas} 
optimal performance is generally achieved only for sufficiently large matrices. 
The symmetry-induced block structure, however, replaces a single large contraction by multiple contractions over smaller tensors, which reduces the efficiency of the BLAS routines and introduces a computational overhead. 
For larger systems, this effect is partially mitigated as the individual symmetry blocks themselves grow in size, thereby improving the efficiency of the underlying matrix multiplications. 
In addition, a more even distribution of tensor indices among the IRREPs enhances the effective reduction in computational cost. 
Preliminary calculations indicate that, for typical quantum-chemical applications up to the CCSDT truncation level, substantial computational savings are obtained for $n_G\leq 8$ when exploiting complex Abelian PGs.

\section{Implementation}
The exploitation of complex Abelian PGs of order $n_G\leq 8$ in the context of finite-magnetic field calculations has been implemented in the \textsc{cfour}\cite{cfour,Matthews_2020} and \textsc{qcumbre}\cite{Hampe2019, qcumbre} program packages. The implemented PGs are $C_{n}$, with $n=3-8$, $C_{nh}$ with $n=3,4$, and $S_{n}$ with $n=4,6,8$.

The DCD as described in sec.~\ref{sec:DCD} is  utilized in the \textsc{mint} module\cite{mint} of \textsc{cfour} for the calculation of integrals over London orbitals using the McMurchie-Davidson algorithm.\cite{McMurchie1978,Tellgren2008} 
The implementation is based on a pre-existing implementation by Gauss for the case of real Abelian PGs. 
We note that the use of complex Abelian PGs was also extended for the use with the Cholesky decomposition of the two-electron integrals over London orbitals.\cite{Blaschke2021,Gauss2022,Blaschke2024} 
Thus, SAOs are employed in ff-SCF calculations within \textsc{cfour}. 

Accordingly, the block structure as described in sec.~\ref{sec:tens} has been implemented in the back-end of \textsc{qcumbre}.\cite{qcumbre}  As such, all currently available post-HF approaches within the program are able to exploit both real and complex Abelian PGs. Future implementations within \textsc{qcumbre} hence only need to explicitly consider the IRREP of the involved tensors and operators in order to function with the presented block structure.

Further details on the implementations within \textsc{cfour} and \textsc{qcumbre} can be found in ref.~\onlinecite{Kitsaras2023}.

The validity of the implementation was verified by comparing energies and molecular properties obtained at both the SCF and correlated levels of theory. Results from calculations employing complex and real Abelian PGs were benchmarked against corresponding calculations performed without symmetry exploitation.

\begin{table*}[tb]
\begin{tabular}{c | c l | c | c} 
     System & \multicolumn{2}{c|}{Molecule} &  Magnetic field (PG)& Largest real Abelian PG \\ \hline
    1) &
     \centering  \scalebox{0.6}{
     \begin{tikzpicture}[baseline=(current bounding box.center)]
          \draw[->] (0.7,0.3) -- +(0,0.5);
          \node at(0,0){\chemfig{C([:90,1]-H)([:-19,1]-H)([:240,1]<H)([:200,1]<:H)}};
    \end{tikzpicture}
    }

    &  Methane  & Parallel to a C-H bond ($C_3$) & $C_1$ \\
    2) &
    \centering  \scalebox{0.6}{\begin{tikzpicture}[baseline=(current bounding box.center)]
            \draw  (0.75,0.75) circle (0.15);
            \draw[fill] (0.75,0.75) circle (0.015);
          \node at(0,0){\chemfig{C([:90,1]-H)([:-180,1]-H)([:0,1]-H)([:-90,1]-H)}};
    \end{tikzpicture}} 

    & Planar methane & Perpendicular to the plane ($C_{4h}$) & $C_{2h}$ \\
    3) &
    \centering  \scalebox{0.6}{
    \begin{tikzpicture}[baseline=(current bounding box.center)]
          \draw[->] (-0.25,0.7) -- +(0.5,0);
          \node at(0,0){\chemfig[scale=0.4]{C([:120,1]-H)([:-140,1]<H)([:-100,1]<:H)([:0,1]-C([:-60,1]-H)([:40,1]<:H)([:80,1]<H))}};
    \end{tikzpicture}
    }

    &Ethane   & Parallel to the C-C bond ($S_{6}$) & $C_{i}$ \\
    4) &
    \centering  \scalebox{0.6}{
    \begin{tikzpicture}[baseline=(current bounding box.center)]
          \draw[->] (-0.4,0.7) -- +(0.5,0);
          \node at(0,0){\chemfig{C([:120,1]-H)([:-120,1]-H)([:0,1]=C([:0,1]=C([:20,0.8]<:H)([:-20,0.8]<H)))}};
    \end{tikzpicture}
    }

    & Allene  & Parallel to the C-C-C axis ($S_{4}$) & $C_{2}$ 
\end{tabular}
\caption{Molecular systems used to assess computational cost reduction via complex Abelian point-group symmetry.}
\label{tab:mol_timing}
\end{table*}

\section{Computational details}
To demonstrate the computational speedup achieved by exploiting the implemented complex Abelian PGs, quantum-chemical calculations at the ff-CCSD level of theory using London orbitals were performed on four molecular systems. The systems considered are small hydrocarbons in the presence of a magnetic field in an orientation that exhibits a complex Abelian PG symmetry. Specifically, the molecules studied are 
1) \ce{CH4} in a standard tetrahedral configuration (C$_3$), 
2) \ce{CH4} in a planar configuration  (C$_{4h}$), 
3) \ce{CH3-CH3} in a staggered configuration (S$_6$), 
and 4) \ce{CH2=C=CH2} (S$_4$). 
The molecules were chosen mainly for their symmetry properties though they might be of interest for atmospheres of DQ white dwarf stars where \ce{CH} and \ce{C2} bands are have been recorded.\cite{Berdyugina2007,Kowalski2010,Vornanen2013,Kitsaras2024,Kitsaras2025,Grazioli2025} 
Note that while planar methane, system 2), for example, does not correspond to an equilibrium structure of the molecule, methane may drastically deform in the presence of a magnetic field and acquire a planar fan-like geometry for strong magnetic fields.\cite{Pemberton2022}
The employed orientation of the magnetic field in the aformentioned systems are depicted in Tab.~\ref{tab:mol_timing}. 
The magnetic field strength was $B= 0.1\ B_0$ for systems 1)-3) and $B= 0.001\ B_0$ for system 4).

The calculation of integrals and the solution of the HF equations were performed using the \textsc{cfour} program package.\cite{Matthews_2020,cfour,mint} The transformation of integrals from the (S)AO to the molecular-orbital (MO) basis as well as the CCSD calculation were performed with the \textsc{qcumbre} program.\cite{qcumbre} 
The correlation-consistent cc-pV$X$Z basis sets, with $X=\mathrm{D,T,Q}$, were employed.\cite{dunning1989a}  
Calculations were performed using the full complex PG and the largest real Abelian subgroup of each system, as listed in the same table.  The resulting computational times are compared with those obtained from calculations without symmetry exploitation, corresponding to $C_1$ as computational PG. 
To ensure a meaningful comparison of computational timings, all calculations for a given system were executed on a single node using a single thread, with no concurrent jobs running. 
Specifically, calculations for system 1) and 2) were performed on an Intel(R) Xeon(R) CPU E5-2643 v4 \@ 3.40GHz, CPU 12 Core, Memory 1,5 TB, Disk 6 TB architecture, while systems 3) and 4) were performed on an Intel(R) Xeon(R) CPU E5-2699 v4 \@ 2.20GHz, CPU 44 Core, Memory 756 GB, Disk 6 TB architecture.

\section{Results}
The speedup observed for calculations on systems 1)–4) upon exploiting molecular symmetry is reported in Table~\ref{tab:timings}.
The individual steps of a CCSD calculation were analyzed separately. These steps include:~a) the evaluation of integrals over London orbitals, which scales as $\mathcal{O}(N^4)$,~b) the solution of the HF problem using an SCF algorithm scaling as $\mathcal{O}(N^4)$ as well,~c) the transformation of integrals from the (S)AO basis to the molecular orbital (MO) basis, scaling as $\mathcal{O}(N^5)$, and~d) the solution of the CCSD equations, which scales as $\mathcal{O}(N^6)$.
It is worth noting that, at the algorithmic level, only the integral calculation using the DCD approach differs between the complex and real Abelian PG implementations.
The symmetry handling of all other steps is based on a unified implementation for both cases.
To quantify the speedup, execution times from calculations performed in $C_1$ are compared to those obtained when molecular symmetry is exploited, giving the ratio 
\begin{align}
r=\frac{t(C_1)}{t(\textrm{Comp. PG})}.
\end{align}
This ratio can be found in the central columns of Tab.~\ref{tab:timings}. 
To further evaluate the implementation and relate it to the theoretically derived cost reduction discussed in Sec.~\ref{sec:tens}, the ratio is examined in terms of powers of the point-group order $n_G$. 
To assess whether the expected up to quadratic behavior is observed numerically, the exponent $\mathrm{log}_{n_G}(r)$ which is related to the ratio by
\begin{align}
r = n_G^{\mathrm{log}_{n_G}(r)},  
\end{align}
is reported in the columns ext to the ratios 
in Tab.~\ref{tab:timings}. 
Lastly, in the rightmost columns the number of iterations needed to achieve convergence for the SCF and CCSD steps are presented.
For the integral evaluation, the speedup due to the use of the DCD is highly system specific and does not depend on $n_G$ directly, but instead on the number of symmetry equivalent centers, as discussed on the topic of stabilizers in sec.~\ref{sec:DCD}.
The speedup of all other steps  stems mainly from the block structure of the tensors involved and the contractions between them. 
Note that for very short time steps, the timing resolution of the \textsc{qcumbre} calculations did not permit the evaluation of the ratio.

\begin{table*}[tb]
    \centering
    \resizebox{\linewidth}{!}{
    \begin{tabular}{c|c||c|c|c|c|c||c|c|c|c|c|| c | c }
        \multirow{2}{*}{Basis set (\# AOs)} & \multirow{2}{*}{Comp. PG ($n_G$)} &\multicolumn{5}{c||}{Ratio $r$ ($C_1/\mathrm{Comp.\ PG}$)}& \multicolumn{5}{c||}{$\mathrm{log}_{n_G}(r)$} & \multicolumn{2}{c}{\# of iterations}\\
        & &Integrals  & SCF  & AO to MO  & CCSD    & Total  &Integrals & SCF  & AO to MO  & CCSD   & Total &SCF& CCSD   \\  \hline  \hline
         
        \multicolumn{12}{c}{1) Tetrahedral methane \ce{CH4} \ } \\   \hline  
        cc-pVDZ ($34$) &$C_{3}$ ($3$) &  2.2 & 3.0 & -\fnm[1] & 2.0 & 1.5 & 0.72  &1.00& -\fnm[1]&0.63&0.37 & 14 & 10\\ \hline
        cc-pVTZ ($86$)&$C_{3}$ ($3$) &  2.4 & 7.1 & 5.6 & 4.3 & 3.8 & 0.80&1.78&1.57&1.33&1.22  &14  & 10 \\ \hline
        cc-pVQZ ($175$)&$C_{3}$ ($3$) &  2.6 & 6.6 & 5.4 & 3.9 & 3.9 &0.87 & 1.72 & 1.54 & 1.24 & 1.24 &14 & 10 \\ \hline \hline

        \multicolumn{12}{c}{2) Planar methane \ce{CH4}} \\   \hline
        \multirow{2}{*}{cc-pVDZ ($34$)}&$C_{2h}$ ($4$) &  2.1 & 4.0 & -\fnm[1] & 2.0 & 1.5 &0.54 & 1.00 & -\fnm[1] & 0.50 & 0.29 & 13 & 11  \\ 
        &$C_{4h}$ ($8$) &  2.2 & 8.0 & -\fnm[1] & 1.0 & 1.2 & 0.38 & 1.00 & -\fnm[1] & 0.00 & 0.09 & 13 & 11  \\  \hline
        \multirow{2}{*}{cc-pVTZ ($86$)}&$C_{2h}$ ($4$) &  2.4 & 9.2 & 7.3 & 6.3 & 4.7 & 0.63 & 1.60 & 1.43 & 1.33 & 1.12 & 14 & 11 \\
        &$C_{4h}$ ($8$)  & 2.8 & 34.1 & 29.0 & 5.7 & 5.4 & 0.50 & 1.70 & 1.62 & 0.84 & 0.81 & 14  & 11 \\  \hline
        \multirow{2}{*}{cc-pVQZ ($175$)}&$C_{2h}$ ($4$) &  2.5 & 9.7 & 7.2 & 6.0 & 5.0 & 0.66 & 1.64 & 1.42 & 1.29 & 1.16 & 14 & 11 \\ 
        &$C_{4h}$ ($8$)  &  4.1 & 34.0 & 25.6 & 7.9 & 8.9 & 0.68 & 1.70 & 1.56 & 0.99 & 1.05 & 15 & 11 \\ \hline \hline

        \multicolumn{12}{c}{3) Staggered ethane \ce{CH3-CH3} } \\   \hline
        \multirow{2}{*}{cc-pVDZ ($58$)}&$C_{i}$ ($2$)&  2.0 & 3.1 & 2.5 & 3.3 & 2.6 & 1.00 & 1.63 & 1.32 & 1.72 & 1.38 & 13 & 11  \\ 
        &$S_{6}$ ($6$) &  2.6 & 17.0 & -\fnm[1] & 6.6 & 4.5 & 0.53 & 1.58 & -\fnm[1] & 1.05 & 0.84  & 15 & 11 \\ \hline
        \multirow{2}{*}{cc-pVTZ ($144$)}&$C_{i}$ ($2$)&  2.2 & 3.1 & 2.6 & 2.8 & 2.6 & 1.14 & 1.63 & 1.38 & 1.49 & 1.38 & 14 & 11  \\ 
        &$S_{6}$ ($6$)  &  2.3 & 21.2 & 10.8 & 10.9 & 7.6 & 0.46 & 1.70 & 1.33 & 1.33 & 1.13  & 16 & 11 \\ \hline
        \multirow{2}{*}{cc-pVQZ ($290$)}&$C_{i}$ ($2$)&  2.9 & 3.3 & 3.7 & 2.7 & 3.0 & 1.54 & 1.72 & 1.89 & 1.43 & 1.58 & 14 & 11 \\ 
        &$S_{6}$ ($6$)  &  3.7 & 23.3 & 16.0 & 8.9 & 8.7 & 0.73 & 1.76 & 1.55 & 1.22 & 1.21 & 16 & 11 \\ \hline \hline

        \multicolumn{12}{c}{4) Allene \ce{CH2=C=CH2} } \\   \hline
        \multirow{2}{*}{cc-pVDZ ($62$)}&$C_{2}$ ($2$)  &  1.7 & 3.0 & 2.3 & 3.6 & 3.0  & 0.77 & 1.58 & 1.20 & 1.85 & 1.58  & 17 & 13\\ 
        &$S_{4}$ ($4$) &3.0 &  11.0 & 7.0 & 7.1 & 5.8  & 0.79  & 1.73 & 1.40 & 1.41 & 1.27 & 18 & 13  \\ \hline
        \multirow{2}{*}{cc-pVTZ ($146$)}&$C_{2}$ ($2$) & 1.7 & 3.1 & 2.8 & 2.9 & 2.7  & 0.77 & 1.63 & 1.49 & 1.54 & 1.43 & 17 & 13 \\ 
        &$S_{4}$ ($4$) &  3.5 & 11.1 & 6.5 & 7.8 & 6.7  & 0.90 & 1.74 & 1.35 & 1.48 & 1.37  & 18 & 13\\ \hline
        \multirow{2}{*}{cc-pVQZ ($290$)}&$C_{2}$ ($2$) &  1.8 & 3.1 & 3.4 & 2.7 & 2.7  & 0.85 & 1.63 & 1.77 & 1.43 & 1.43  & 17 & 13\\ 
        &$S_{4}$ ($4$) &  4.3 & 11.6 & 8.7 & 6.6 & 6.5 & 1.05 & 1.77 & 1.56 & 1.36 & 1.35 & 18 & 13 \\ 
        
    \end{tabular}
    }
    \fnt[1]{For very short time steps, the timing resolution of the \textsc{qcumbre} calculations did not permit the evaluation of the ratio.}

    \caption{Reduction in the time needed to complete a calculation at the CCSD level for systems 1)-4).}
    \label{tab:timings}
\end{table*}

It is observed that the cost reduction is greater for larger systems. This expected trend is explained by the fact that 
as the system size increases, a more balanced distribution of matrix indices among IRREPs enhances the computational efficiency achieved by excluding vanishing blocks. Moreover, the employed BLAS\cite{blas} routines in this work perform better for larger matrices.

The integral calculation shows the least speedup compared to the other computational steps.
For the existing implementation employing real Abelian PGs,\cite{mint} the largest speedup for the integral evaluation is observed for system 3) with $n_G^{1.5}$ employing the  $C_i$ PG.  
The lowest speedup for this step is $n_G^{0.5}$. 
On average, for the integral evaluation, a value of $n_G^{0.9}$ is obtained. 
The newly implemented complex Abelian PGs underperform in comparison. 
This reduced efficiency of complex Abelian PGs arises from the fact that 
for the real Abelian PG case, the final DCD expressions involve simple parity factors, cf. Eqs.~\eqref{eq:oneDCR_nosym_re} and \eqref{eq:twoint_final_re}. For complex Abelian PG, on the other hand, the sums over basis functions arising from Eq.~\eqref{eq:rot_AO} 
cannot be simplified further, cf. Eqs.~\eqref{eq:oneDCR_nosym} and \eqref{eq:twoint_final}. 
However, it should be noted that the PGs $S_4$ and $C_{4h}$ are special cases where Eq.~\eqref{eq:rot_AO} has only one single contribution as well, and hence a better performance compared to other complex groups is expected. Excluding these cases, a maximum value $n_G^{0.9}$ is observed for system 1) computed within $C_3$. The poorest performance is observed  for system 2) calculated within $C_{4h}$ using the cc-pVDZ basis set with $n_G^{0.4}$. 
The average of the exponent amounts to $n_G^{0.7}$ for the complex Abelian PGs for the integral evaluation step.

At the SCF level of theory, the complex Abelian PGs slightly outperform the real ones. 
For both cases, the average of the exponents is $n_G^{1.6}$, with maximum values of $n_G^{1.7}$ and $n_G^{1.8}$, respectively. 
The least reduction is in both cases $n_G^{1.0}$. This mildly better performance of the complex Abelian groups is observed even for the cases where up to two more iterations are needed in the case of complex Abelian PGs, as seen in Tab.~\ref{tab:timings}.

Both the integral transformation and the CCSD iterations are performed within \textsc{qcumbre}. For the integral transformation, an average speedup of approximately $n_G^{1.5}$ is observed for both complex and real PGs, with a slightly better performance for the complex case.
For the CCSD iterations, more pronounced differences are found. The average speedup amounts to $n_G^{1.1}$ for complex PGs and $n_G^{1.4}$ for real PGs.
Although this difference appears substantial, it is important to note that the calculation is much more efficient when using the complex PG even if the exponent of the ratio in the base of $n_G$ is smaller. 
To explain this difference and the deviation from the quadratic speedup, we note that in the implementation used, tensor contractions are performed using BLAS matrix multiplication routines.\cite{blas} As discussed in sec.~\ref{sec:tens}, these multiplication routines exhibit optimal performance for larger matrices and, as such,  are less efficient for larger groups in a given system.
Furthermore, we note that, in the examples discussed here, the occupied orbitals are distributed rather inhomogeneously among the IRREPs for the complex Abelian PGs, which leads to a reduced efficiency as well.
Because both real and complex PGs are treated within the same implementation and the number of CCSD iterations does not differ, the observed difference is attributed to the lower order of the real PGs rather than to an intrinsic disadvantage of complex Abelian PGs. For comparable group orders $n_G$, or for larger systems, the aforementioned differences should diminish.

The overall speedup observed reflects the reduction in computational effort associated with solving the CCSD amplitude equations, which constitute the computationally most demanding part of the procedure. The average scaling exponents obtained are $n_G^{1.0}$ and $n_G^{1.3}$ for complex and real PGs, respectively. 
While the ideal theoretical scaling of 
$n_G^2$ is not fully reached, the observed behavior remains substantial. Deviations from the asymptotic limit are largely due to the small sizes of individual tensor blocks, which reduce the efficiency of matrix multiplication routines. Overall, the measured speedups demonstrate that symmetry exploitation provides significant and practically relevant performance gains.

It is also worth mentioning that, in an absolute sense, the complex Abelian PGs always outperform the use of the largest Abelian subgroup as seen by the calculated ratios.

Finally, we note that the exploitation of symmetry not only reduces computational cost, but it further facilitates 
the analysis of orbitals and excited states thereby acting as an interpretational tool, and 
enables the selective targeting of states belonging to specific IRREPs. 
Beyond ff-CCSD, we have performed calculations exploiting complex PGs 
for excited states using the EOM-CC ansatz.\cite{Kitsaras2023,Kitsaras2024,Blaschke2024,Kitsaras2025,Grazioli2025} In these studies, the use of the aforementioned groups and identification of the IRREPs of the electronic states involved played a crucial role. 
We mention for example the study of \ce{B(OH)3} in Ref.~\onlinecite{Grazioli2025}, a molecule that exhibits a complex PG even in the absence of a magnetic field.
Here, the HOMO-LUMO transition belongs to IRREPs with complex characters and is \emph{accidentally} doubly degenerate in the field-free case. As such, targeting these states at the EOM-CCSD level is impossible when using a real implementation. Using the complex Abelian PGs with our complex code facilitated their targeting and analysis.

\section{Conclusion}
\label{sec:conclusion}
In this work, an implementation of complex Abelian PGs for quantum-chemical calculations in the context of finite-magnetic-field approaches was presented. 
The theoretical aspects presented cover the totality of an electronic-structure treatment, from the calculation of integrals over atom-centered basis functions to considerations for SCF and post-HF methodologies.

The implemented approaches were applied to small hydrocarbons in the presence of a finite magnetic field. 
Calculations at the ff-CCSD level were carried out to illustrate the associated decrease in computational cost. 
It can be concluded that, for a quantum-chemical code employing complex algebra in the context of finite magnetic fields, the use of complex Abelian PGs consistently provides significant efficiency gains over real Abelian subgroups, owing to the larger number of symmetry elements typically encountered as compared to the real Abelian subgroups. 
While the implementation presented here is limited to the case of finite-magnetic field approaches, similar gains are expected to be observed for other cases where complex wavefunctions occur. 
\acknowledgements
Funding for this work was provided by the Deutsche Forschungsgemeinschaft via grant STO-1239/1-1.
The authors thank Prof. Jürgen Gauss for providing the original code on which the present implementation is based, as well as for helpful feedback on the manuscript.
The authors furthermore thank Prof. Peter R. Taylor for valuable discussions. 


\section*{Data availability statement}
The data that support the findings of this study are available within the article. 


\section*{References}

\bibliography{biblio}

\end{document}